# Analyzing and Visualizing the Semantic Coverage of Wikipedia and Its Authors

**Todd Holloway**
Indiana University
Department of Computer Science
150 S. Woodlawn Ave.
Lindley Hall 215
Bloomington, IN 47405, USA
Phone: (812) 219-2815
Email: tohollow@cs.indiana.edu

**Miran Božičević**
Wikipedia Networks Team
Multimedia Institute (http://www.mi2.hr)
Preradovićeva 18
HR-10000 Zagreb
Croatia
Email: miran@mi2.hr

**Katy Börner***
Indiana University, SLIS
10th Street & Jordan Avenue
Main Library 019
Bloomington, IN 47405, USA
Phone:  (812) 855-3256   Fax: -6166
E-mail: katy@indiana.edu
WWW: http://ella.slis.indiana.edu/~katy

* To whom all correspondence and proofs are to be addressed.

 **Number of text pages: 20**

**Number of figures: 6**

**Number of tables: 3**

**Abstract**

This paper presents a novel analysis and visualization of English Wikipedia data. Our specific interest is the analysis of basic statistics, the identification of the semantic structure and age of the categories in this free online encyclopedia, and the content coverage of its highly productive authors.

The paper starts with an introduction of Wikipedia and a review of related work. We then introduce a suite of measures and approaches to analyze and map the semantic structure of Wikipedia. The results show that co-occurrences of categories within individual articles have a power-law distribution, and when mapped reveal the nicely clustered semantic structure of Wikipedia. The results also reveal the content coverage of the article's authors, although the roles these authors play are as varied as the authors themselves. We conclude with a discussion of major results and planned future work.

**Summary of results for the nonspecialist:**

Wikipedia is a free 'encyclopedia of everything' that was started by Jimmy Wales on January 15[th], 2001. Less than five years after its creation it comprises over 2,700,000 articles written by about 90,000 different contributors in 195 languages. This paper provides basic statistics, analyzes and maps the semantic structure of the English Wikipedia as well as the activity of its major authors.

## 1. Introduction

Prior research[1,2] has shown that particular areas of science are not driven by single authors but by effectively collaborating co-authorship teams – a global brain seems to be emerging on this planet[1]. This has been interpreted as good news as human brains are assumed to not scale to process, understand and manage the amounts of information and knowledge available today. However, teams might be able to dynamically respond to the increasing demands on information processing and knowledge management. In this paper we study one of the most surprising team efforts: the wiki based online Wikipedia. Subsequently, we introduce Wiki technology and the Wikipedia effort, data accuracy and existing biases, data licensing issues, and the Wikipedia community.

### *1.1 Wiki Technology and Wikipedia*

The 'wiki' technology (http://c2.com/cgi/wiki?WikiWikiWeb) was invented by Ward Cunningham in 1995[3]. The defining feature of wikis is that each page has a 'edit this page' link that takes users to an editing view of the page's content. A user can make and submit changes to the text, which immediately replaces the previous version of the text. Hence, readers can easily become authors of the page. Users can register to create and retain a user profile or decide to remain anonymous. When anonymous users make changes, their IP addresses are logged. Each wiki page also has a 'page history' link that provides access to previous versions of the page, as well as a 'recent changes' link that lists most recent edits and helps track changes.

The largest public wiki is Wikipedia, a free 'encyclopedia of everything', that was started by Jimmy Wales on January 15th, 2001. Less than five years after its creation it comprises over 2,700,000 articles written by about 90,000 different contributors in 195 languages. Three racks of servers process 60 million Wikipedia hits per day, serving the information needs of more users than, e.g., britannica.com or nytimes.com[4].

Wikipedias in different languages are only loosely interlinked. There were nine different Wikipedias that contained more than 50,000 articles as of Nov. 5, 2005—German, English, French, Italian, Japanese, Dutch, Polish, Portuguese, and Swedish. The largest Wikipedia is in English and comprises 800,342 articles and 78,977 categories. Users may add links between pages in different language Wikipedias, but otherwise the Wikipedias exist independent of one another.

|                                           | **English Wikipedia** | **All Wikipedias** |
|-------------------------------------------|----------------------:|-------------------:|
| Number of articles (official Wikipedia count) | 800,342           | 2,740,886          |
| New articles per day                      | 1,515                 | 5,654              |
| Number of internal links                  | 16.8 million          | 45.3 million       |
| Total number of words                     | 292 million           | 735 million        |
| Mean size of article (bytes)              | 2,729                 | 2,478              |
| Mean number of edits per article          | 23.4                  | 16.2               |
| Total contributors                        | 43,531                | 89,921             |
| Active contributors (> 5 edits)           | 14,434                | 28,258             |
| Very active contributors (> 200 edits)    | 1,854                 | 4,573              |
| New contributors                          | 2,062                 | 4,274              |

**Table 1.** Wikipedia statistics. All data from Wikipedia Statistics[5] for October 2005, except number of articles, from Wikipedia:Multilingual Statistics[6] for November 1, 2005.

*1.2 Accuracy, Bias and Persistence*
Wikipedia's reliability as a source of information has been repeatedly questioned in the media[7,8] and debated in its communities. Critics see two main flaws: persistent inaccuracies and systemic bias.

The inaccuracies stem from the lack of an authority or a peer review process that would verify every piece of information entered into Wikipedia. Users voluntarily review new edits[9], and while vandalism and self-promotion get identified and reverted quickly, small errors and bad writing may remain on display for significant periods[8]. In early July 2005, for example, an erroneous report of the death of comic strip author Jeph Jacques remained online for two days, surviving through 10 edits.

As a result, Wikipedia is most often not admitted as a reliable bibliographic reference. The English Wikipedia community is working on procedures that would help create a parallel editorially validated version, names Wikipedia 1.0.[10]

A systemic bias arises because English Wikipedia authors are necessarily Internet users with decent knowledge of the language, enough free time to contribute, and sufficient technical facility to edit a wiki page. Wikipedia coverage tends to favor topics of interest to such users. In October 2004, for example, the article about Hurricane Frances was five times the size of one on Chinese art[7]. Wikipedia itself reports imbalanced coverage by geographic area favoring North America, Japan, Western Europe, Australia and New Zealand, significantly more pronounced than in editorially created English language encyclopedias[11]. Wikipedians concede this imbalance as inevitable, with hope it will decrease as the content grows[9].

A third issue is the non-persistence of Wikipedia entries due to continuous update. This makes the value of citing Wikipedia entries questionable.

*1.3 Licensing for Collaboration*

Wikipedia licenses its text under the GNU Free Documentation License (GFDL). Free licenses, first intended for software, appeared in 1984, allowing anyone to use computer applications, inspect and modify their source code, and distribute them without limitation. Introduced on ethical grounds, with belief that 'people should be free to use software in all the ways that are socially useful'[12], they have since shown practical utility, as opening source code made software development more efficient[13].

Interestingly, many top Open Source projects have decided not to use GFDL as it contains provisions for *invariant sections* that can't be modified once published, even if they are inaccurate or contain plagiarized information. Another provision is that such published information cannot be removed, except by its creator. Wikipedia, by adopting the GFDL, is stuck publishing such information despite its veracity/provenance.

Wikipedia is by far the largest online effort that uses such a license. GDFL abolishes individual authorship of articles, leveling the playing field for all contributors, and helps create a sense of shared content ownership by the community[9]. Wikipedians feel that 'authorship data is irrelevant and sometimes even detrimental to the creation of truly communal repositories of knowledge', see [14] and [15]. However, authorship information does seem to add context to interactions and signing is used extensively on talk pages. Author info is also valuable when browsing the recent changes or history pages. Contributions of unknown authors are closely scrutinized.

By espousing an inclusive point of view policy and involving non-experts in the discussion, Wikipedia arguably has the potential to provide an open and dynamic platform complementary to the scientific peer-review process for reasoned debate on issues for which there is no accepted expert view[16].

Finally, a free license helped popularize Wikipedia, as any site can mirror its text without special permission, and serves as an escape route should the work of the Wikimedia Foundation be compromised. Anyone has the freedom to take the contents of Wikipedia and fork the entire project[9]. The Spanish Wikipedia in fact forked in 2002 to found Enciclopedia *Libre Universal en Español*, which still exists independently.

*1.4 Wikipedia Community*

There appears to be a major social incentive for contributing time to this unique effort: Wikipedians form a tightly knit community. They watch over each other and the content they create. Thousands of Wikipedia editors, which Wales characterizes as 'extremely information hungry, geeky kinds of persons',[4] welcome new contributors, help resolve conflicts, enforce policies, etc.

Wikipedia is by no means the first website that relies on massive user participation: anyone can post news on Slashdot, offer goods at eBay or review books at Amazon. These sites, though, facilitate participation through hard-coded reputation mechanisms. Users grade others' contributions, and the website compiles an overall score to help direct further interaction. Reputation is computed based on individual assessments using a predefined algorithm.

Wikipedia, on the other hand, relies on facilitating human interaction rather than superseding it. Encyclopedic content is so complex that 'a process of reasoned discourse' is the only practical way to reach agreement; contributors can get to know each other and a community forms. The decisions on new structures and procedures, such as how to go about deleting articles or when to temporarily block editing, are then delegated to the community as well rather than instituted centrally. Individual reputation forms as 'a natural outgrowth of human interaction'[9].

The resulting constitution of decision making in Wikipedia is hybrid. Members actively avoid majority voting, instead striving to reach *consensus* on any issue, but can use polls (*democracy*) as a non-binding tool in this process. Individual users who gain reputation through their contributions form a merit-based *aristocracy*, with several layers of privilege: anonymous users, regular users, administrators who can, e.g., delete or block pages in a single Wikipedia, and two higher levels that can, e.g., confer administrator status. Mediation and arbitration committees resolve disputes, while a rare issue may require the judgment of the 'benevolent dictator', Mr. Wales (*monarchy*)[9,17].

## 2. Related Work

Prior analyses of Wikipedia data aimed to create timelines of the number of articles, authors, and other elements over time. Erik Zachte gives a detailed recipe for generating such graphical timelines from a simple script[18].

History Flow visualizations[14] show the relationships between multiple document versions and reveal complex patterns of cooperation and conflict. They aims to make broad trends in revision histories visible while preserving details for closer examination. The authors also proposed several hypotheses of how and why Wikipedia succeeds to create high quality content. They pointed out that constant change leading to frequent vandalism and inaccuracy counter acted by rapid and effective repairs is at the center of Wikipedia. They believed that rapid change might be critical for other online communities where collaboration and consensus is critical.

Visually, History Flow diagrams are similar to Theme River[19] and parallel coordinate systems[20] however a different type of data is displayed and the vertical axes are used differently. Using the tool, the authors identified diverse patterns of collaboration and negotiation such as vandalism and repair, anonymity versus named authorship, negotiation and content stability.

The Wikipedia Networks Team from Zagreb, Croatia, is presently compiling a comprehensive analysis of complex networks of article interlinkage in different language Wikipedias. The analysis includes all basic statistical variables of complex networks and examines their trends over Wikipedias of different sizes[21]. Future work will include influence of communities on article structure, modeling growth and studying categorization across languages.

The subsequently reported study differs from existing work in that it aims to analyze and visualize the semantic coverage of Wikipedia and its authors.

## 3. Analyzing Wikipedia

This section details the Wikipedia data format and reports baseline statistics of the English Wikipedia data.

### *3.1 Data Format and Definitions*

The Wikimedia Foundation Inc. generously makes public all current articles and past revisions, providing a rich record of the structure and evolution of Wikipedia content as well as of the activities and roles of its many contributors.

A complete dump of Wikipedia in all languages is freely available from http://download.wikimedia.org. The most recent dump was generated on Nov 5th and is used subsequently.

There are two tables of interest for this study:

- Cur → current contents of English Wikipedia (29,208 MB)

- Categorylinks table → article to category and subcategory to supercategory membership relations (178 MB)

The dumps are generated in mysql and, more recently, in XML formats. To utilize the dumps, the XML containing the cur table was downloaded and loaded into a mysql database using the Java-based tool MWDumper[22]. The categorylinks table is downloadable as compressed SQL and can be loaded directly into the database without the assistance of specialized tools.

All pages in Wikipedia belong to a namespace. This namespace is part of the URL for a given page. Categories belong to the 'Category' namespace; therefore, the URL for the category 'Information Science' is http://en.wikipedia.org/wiki/Category:Information_science. In the Wikipedia world, a category is a 'list page' which serves to classify other pages. Categories exist in a format identical to article pages except that the subcategory links are not explicitly stored, but generated from the text of child pages. Articles belong to the main namespace, meaning that they do not require a prefix in the URL. In this paper, the term article refers to any page in the main namespace. The Wikipedia definition of an article[23] is more narrow, as stub, disambiguation, redirect, and other types of pages are in the main namespace but are not considered articles. Thus we have 1,553,648 articles, whereas the official article count is 800,342. We use this broader definition because we are interested in the organization of categories and pages pointing to categories, and all pages in the main namespace play a role in this analysis.

### 3.2 Statistics

This section provides baseline statistics for the article and category data extracted from the Nov 5th, 2005 data dump. This cur dump contains 1,553,648 unique articles, 78,977 unique categories, and 39,598 unique authors. Figure 1 shows the number of articles, categories, and contributors (last edit) in the cur table over time. This figure shows that Wikipedia is broadly updated. The introduction of categories in May 2004 is clearly visible.

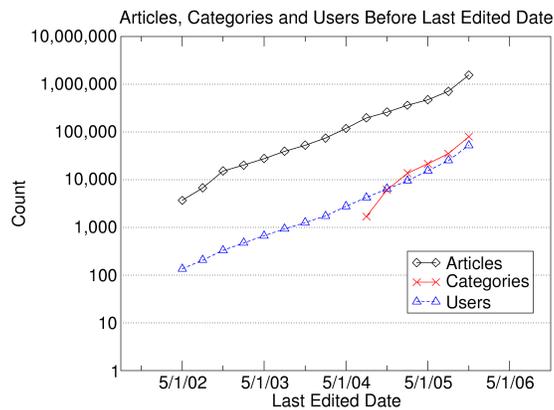

**Figure 1**: Number of articles, categories, and contributors (last edit) over time.

Figure 2 shows the distribution of categories per article and the distribution of articles per category. Out of the 1,553,648 unique articles, 785,858 articles, or roughly half, have no category assignments. The articles "List of publications in biology" and "Chemistry resources" have the most categories with 39 and 44, respectively. Only 2,780 articles have more than 10 categories assigned.

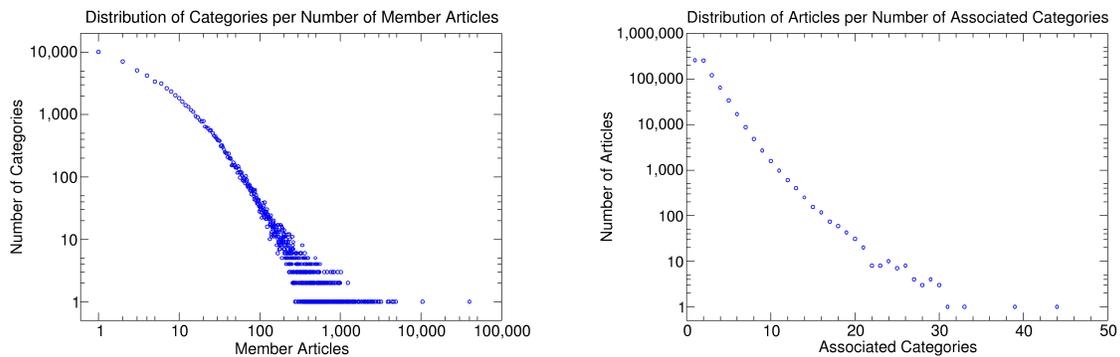

**Figure 2:** Distributions of the number of categories per number of member articles (left) and the number of articles per number of associated categories (right). Neither distribution exhibits clear-cut power law behavior. A consistent exponent cannot be found for the distribution of categories, while the distribution of articles is closer to exponential.

Out of the 78,977 categories, 12,252 are not assigned to any article and 10,116 are assigned to exactly one article. On average 1.17 categories are assigned per article, 2.39 among those articles having at least one category. The top 20 most often used categories are listed in Table 2. The table also shows that stub markers (indicated by *_stub) and other templates play a major role in the categorization process. Templates are custom tags that contain standardized text to be added to an article, such as "This article is a stub". They often automatically create a membership relationship between an article and a specific category. Out of the top 20 most frequently used categories shown, only three ("American actors", "Film actors", "Television actors") do not result from the use of templates.

| Top Categories | # Articles | | |
|---|---|---|---|
| 1. Disambiguation | 40,062 | 11. Politician_stubs | 3,021 |
| 2. 1911_Britannica | 10,450 | 12. Articles_to_be_merged | 2,899 |
| 3. Film_stubs | 4,867 | 13. British_people_stubs | 2,706 |
| 4. Musician_stubs | 4,575 | 14. American_politician_stubs | 2,694 |
| 5. American_actors | 4,551 | 15. Television_stubs | 2,540 |
| 6. American_people_stubs | 4,401 | 16. American_actor_stubs | 2,492 |
| 7. Film_actors | 4,023 | 17. Cleanup_from_October_2005 | 2,466 |
| 8. Musical_group_stubs | 3,873 | 18. Incomplete_lists | 2,362 |
| 9. Album_stubs | 3,859 | 19. Football_(soccer)_biography_stubs | 2,298 |
| 10. Television_actors | 3,207 | 20. Medicine_stubs | 2,297 |

**Table 2:** Top 20 most frequently used categories.

From the 39,598 unique authors, 17,287 were registered and added or modified exactly one article. On average, each registered author added or modified 33.8 articles. The number of articles that were last modified by an unregistered user is 214,594.

**4. Analyzing and Mapping Wikipedia Data**

This section details the generation of a category base map based on the co-occurrence of categories in articles. We then use this base map to map the position of major semantic topics, the last edit time of articles, and the topic coverage of major Wikipedia authors in this semantic space.

*4.1 Category Map Generation*

The semantic space of topics covered by Wikipedia can be mapped based on articles or based on topics. Text analysis and/or linkage analysis techniques could be employed. For the present study, we decided to generate a base map using categories and a measure of similarity of categories.

As mentioned before, category assignments were introduced in May 2004 and 78,977 unique categories have come into existence since then. Categories are organized in a semi-hierarchical fashion, were semi means that there is a root category, called 'Categories', from which most categories can be reached by traversing the subcategories. There are, however, cycles within this category structure, i.e., category A might be a subcategory of category B, B might be a subcategory of C, and C be a subcategory of A. An article might point to a category and its subcategory. There are 1,069 categories that are disconnected from the large component rooted at 'categories'.

| Wikipedia | | Britannica | | Encarta | |
|---|---|---|---|---|---|
| Top Level Categories | 2nd Level Categories | Top Level Categories | 2nd Level Categories | Top Level Categories | 2nd Level Categories |
| Culture | Academia<br>Archaeology<br>Architecture<br>Archives<br>+69 more | Arts & Literature | Major branches<br>Regional and cultural traditions<br>Styles and movements | Art, Language, and Literature | Nat. & reg. literature<br>Literature & writing<br>Architecture<br>Artists<br>+9 more |
| Fundamental | Communication<br>Knowledge<br>Nature<br>Systems<br>Thought | The Earth & Geography | Earth<br>Earth, geological history of<br>The Earth's crust<br>Processes<br>+7 more | Life Science | Plants<br>People in Life Science<br>Medicine<br>Invertebrate Animals<br>+10 more |
| Geography | Africa<br>Americas<br>Antarctica<br>Arctic<br>+42 more | Health & Medicine | Human body<br>Human life<br>Health and disease<br>Medicine and disease<br>Human psychology | History | History of Asia & Australasia<br>People in Euro. History<br>People in US History<br>+6 more |
| History | Archaeology<br>History_books<br>Fictional_history<br>Historical_documents<br>+16 more | Philosophy & Religion | Philosophy<br>Religion | Geography | World cities, towns, …<br>Regions of the World<br>Rivers, Lakes, …<br>+10 more |
| Mathematics | Algebra<br>Mathematical_analysis<br>Applied_mathematics<br>Arithmetic<br>+62 more | Sports & Recreation | Sports<br>Hobbies and games<br>Physical and outdoor recreation | Religion & Philosophy | Theology & Practices<br>Mythology<br>Religious Figures<br>Philosophy<br>+3 more |
| People | People in Positions of Authority<br>Autodidacts<br>Biblical people<br>+2 more | Science & Mathematics | Biological sciences<br>Earth sciences<br>Mathematics<br>Physical Science<br>+2 more | Physical Science & Technology | Construction & Engineering<br>Chemistry<br>Earth Science<br>+13 more |
| Portals | Computer_<br>and_video_<br>games<br>Cricket<br>+2 more | Life | Biology<br>Fossil<br>Life-support system<br>Living things<br>+4 more | Social Science | Economics & Business Organizations<br>Institutions<br>Political Science<br>+8 more |
| Science | Applied sciences<br>Astronomy<br>Biology<br>Science_books<br>+36 more | Society | Civilization<br>Culture<br>Social behavior<br>Social change<br>+7 more | Sports, Hobbies, & Pets | Sports<br>Sports Figures<br>Games, Hobbies, & Recreation<br>Pets |
| Society | Computing_and_society<br>Disability<br>Human_Societies<br>People<br>+7 more | Technology | Industry<br>Material<br>Power<br>Research and development<br>+14 more | Performing Arts | Theater<br>Musicians & Composers<br>Cinema, Television, &<br>+3 more |
| Categories By Topic | Categories_by_continent<br>Categories_by_country<br>Films_by_topic<br>Genres<br>+4 more | History | Archaeology<br>Nations<br>Major eras<br>Study of history | | |
| Topic Lists | Lists_of_country-related_topics<br>Mathematical_lists | | | | |
| Wikiportals | Culture<br>Geography<br>History<br>Human_Societies<br>+4 more | | | | |

**Table 3:** Category hierarchies for Wikipedia (left) Britannica.com (middle) and Microsoft Encarta.com (right)

Traversing the category hierarchy, beginning at 'Categories', to a depth of three, results in the hierarchy shown in Table 3. This hierarchy is contrasted with the top two hierarchy levels of Britannica.com and Encarta.com. Both were read out on Nov. 5th, 2005. The free online versions of Britannica and Encarta have 120,000 and 4,500 articles respectively. Matching category names are color coded.

Given the quality of category interlinkages, we decided to define the similarity of categories based on their co-occurrence in articles. That is, two categories are assumed to be similar to each other and are connected by a link if they are used in one and the same article.

To indicate the strengths of interlinkage we introduce weights. The weight of a link is derived using a cosine similarity measure frequently used in bibliometric studies[2]. Let $A_k C_i$ indicate a category link between article $A_k$ and category $C_i$ then the cosine similarity $COS_{i,j}$ among two categories $i$ and $j$ is computed via

$$COS_{i,j} = COS_{j,i} = \frac{raw}{\sqrt{\sum_{k=1}^{n} A_k C_i \sum_{k=1}^{n} A_k C_j}}$$

where $raw$ is defined as

$$raw = \sum_{k=1}^{n} \left( A_k C_i + A_k C_j \right).$$

The resulting network has 56,609 category nodes and 2,190,700 weighted co-occurrence links. The non-normalized $raw$ weights have a distribution as shown in Figure 3. The highest $raw$ weight is 2,143 for a link that interconnects the categories 'Film actors' and 'American actors'.

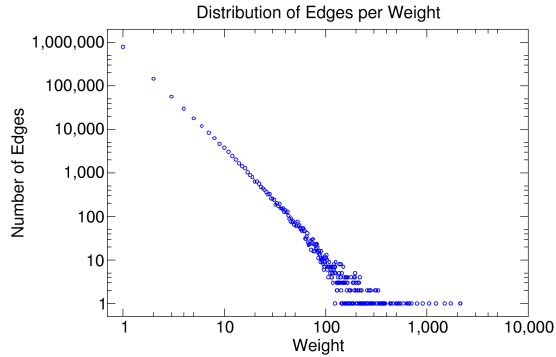

**Figure 3:** Distribution of the number of edges per weight exhibits a power law with exponent γ=-2.96 for weights larger or equal 20.

The resulting weighted category network was loaded into VxInsight[24] a visual-analytic tool for the interactive exploration of large datasets (http://www.cs.sandia.gov/projects/VxInsight/Vxfull.html). VxInsight uses the VxOrd layout algorithm. A 68.6% link cut was applied and 688,456 directed links were retained. The coordinates were saved out and rendered in Pajek[25]. The resulting semantic category base map is shown in Figure 4.

In an attempt to understand the semantic coverage of articles and the semantic interrelations of categories, we used color coding to indicate several major topic areas. Using common category title words, such as 'Companies', 'Death', and 'Film', nodes that contained those words in titles were colored accordingly. In addition, we identified the top level categories previously listed in Table 3. In Figure 4 they are shown as larger, labeled nodes. Four top level categories—'Fundamental', 'Categories by Topic', 'Portals', and 'Wikiportals', are not present in the map as these categories have no pages in the main namespace that link to them.

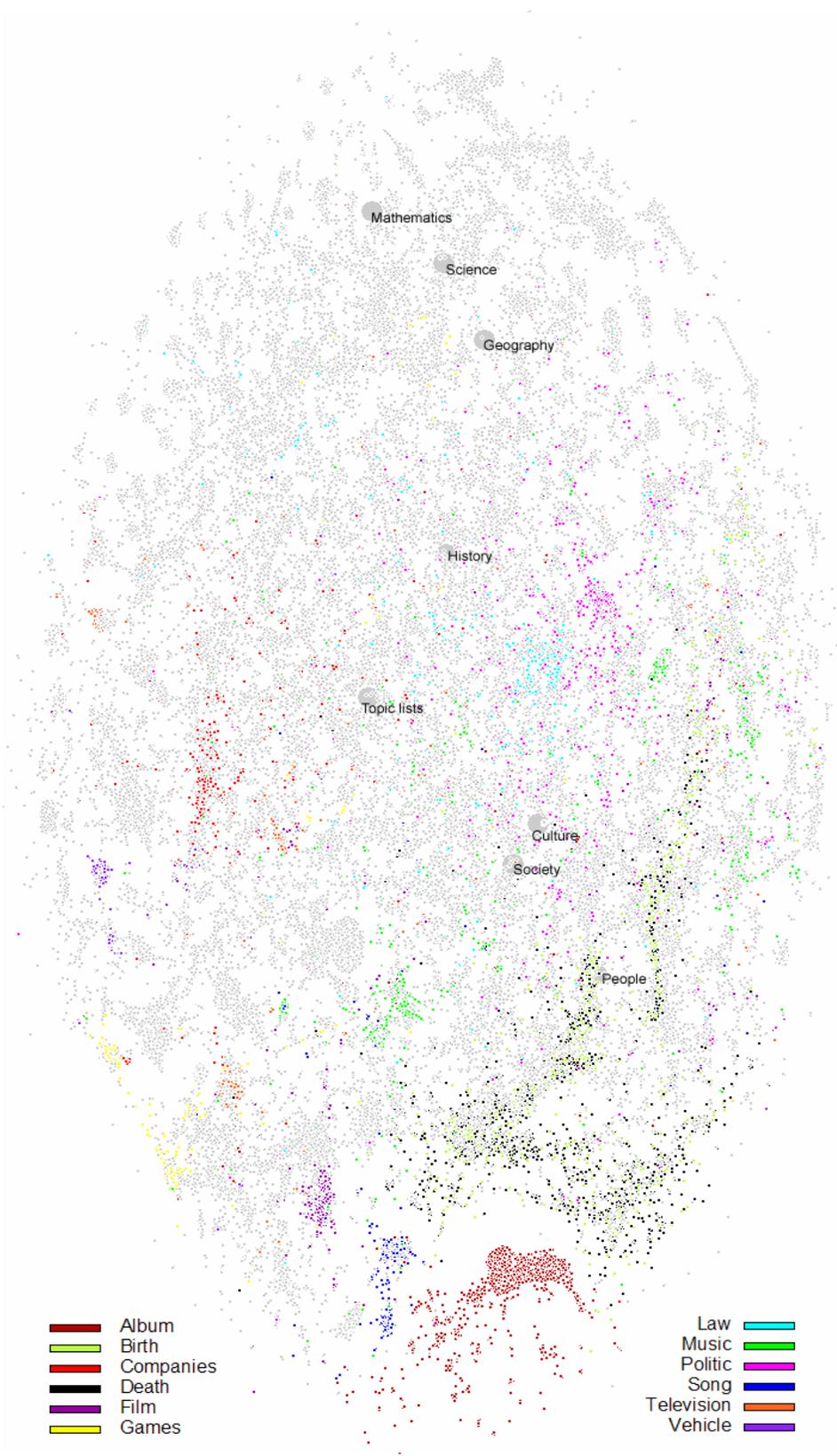

**Figure 4:** English Wikipedia category network laid out in VxOrd and rendered in Pajek.

It is interesting to interactively explore the diverse clusters in this map using VxInsight. Unfortunately, Figure 4 can only present a static snapshot of this unique birds-eye view of the English Wikipedia topic space. However, major category clusters, such as 'Television' related categories in red, 'Song' related categories in blue, or the co-existing 'Death' (black) and 'Birth' (light green), can be easily identified.

*4.2 Mapping Last Edit Time*

In order to see the recentness of categories, the map was color coded by the last edit time given in the Nov $5^{th}$ dataset. Note that edit histories are not taken into account. The result is shown in Figure 5. Old categories are given in black and young categories in light green. This map suggests that category pages are largely kept current, except for some isolated clusters as well as a large region in the upper left consisting of city and county articles clustered by US states. These geographic themed pages were created automatically by the bot 'rambot' from US census data, and thus they may not all have an interested user base yet.

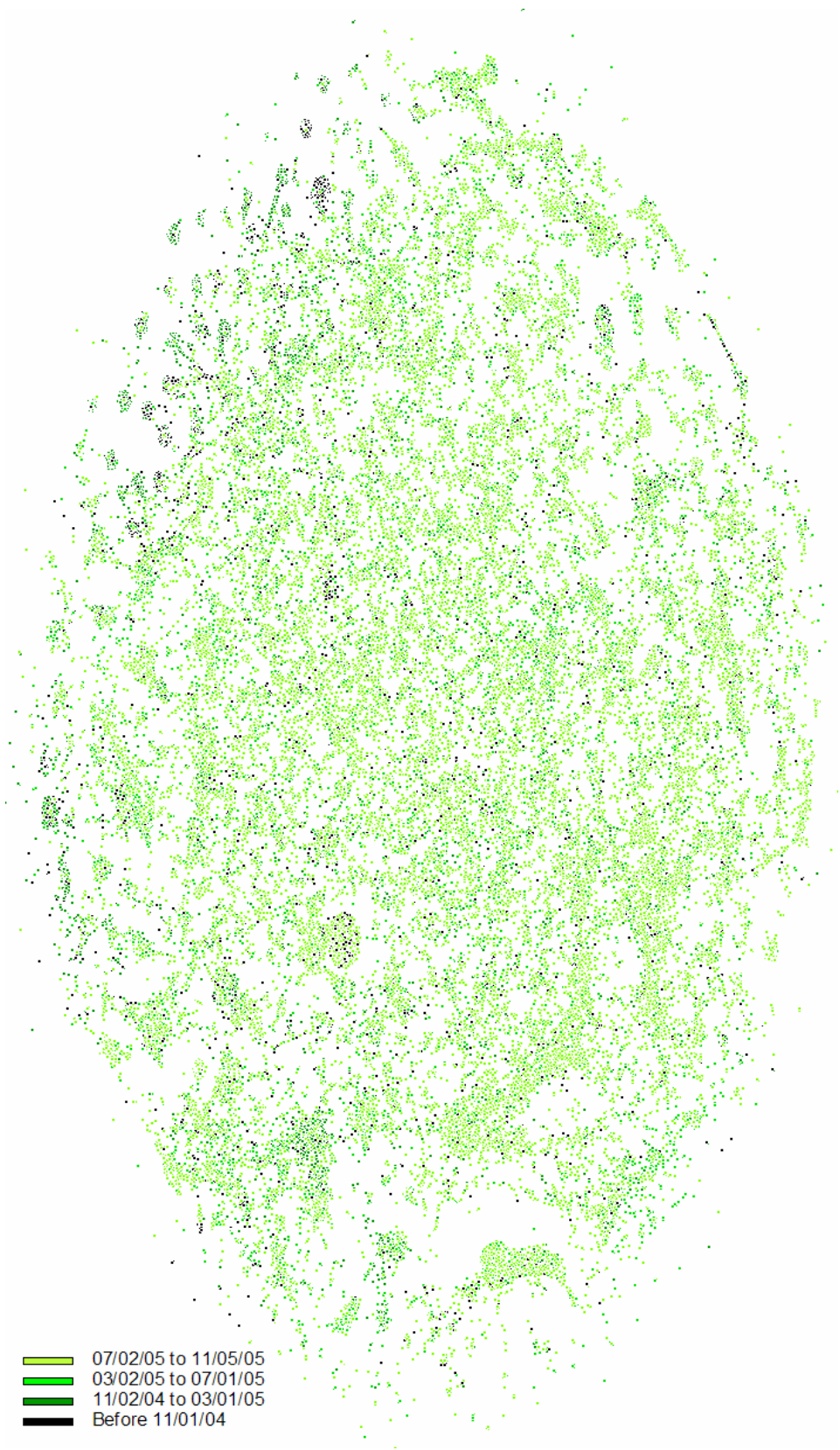

**Figure 5:** English Wikipedia category network color coded by last edit time.

### 4.3 Mapping Topic Coverage of Authors

To understand the topic coverage of individual wiki authors, we plotted the top ten most active category authors. Note that edit histories are not taken into account. A page is exclusively attributed to the author that created or edited a category page last as of November 5$^{th}$, 2005. The result is depicted in Figure 6.

Among the top ten category authors, we find diversity in intentions and scope of category edits. Cross referencing these author names with their user pages on Wikipedia, we find that several are bots, including Whobot (purple), whose primary purpose is categorization, and rambot (blue), whose purpose involves the creation and categorization of pages for cities and counties in the U.S. Among the human authors, we find BDAbramson (light green), an intellectual property lawyer who authors articles primarily about law, but was the most recent author of many categories related to music albums. Rlandmann (black) is a top author of both articles and categories related to aviation. Postdlf (orange) who authored a wide variety of categories explains on his user page his 'categorization philosophy/obsession': 'I don't want to see it done wrong.'

Some of the top ten category authors, including BDAbramson and Whobot, play the administrative role of altering the categories when pages have been deleted, renamed, or merged. Deletion, renaming, and merging occur following nominations for these actions and discussion, and that these individuals implement the results of these discussions greatly increases their presence on a category edits coverage map.

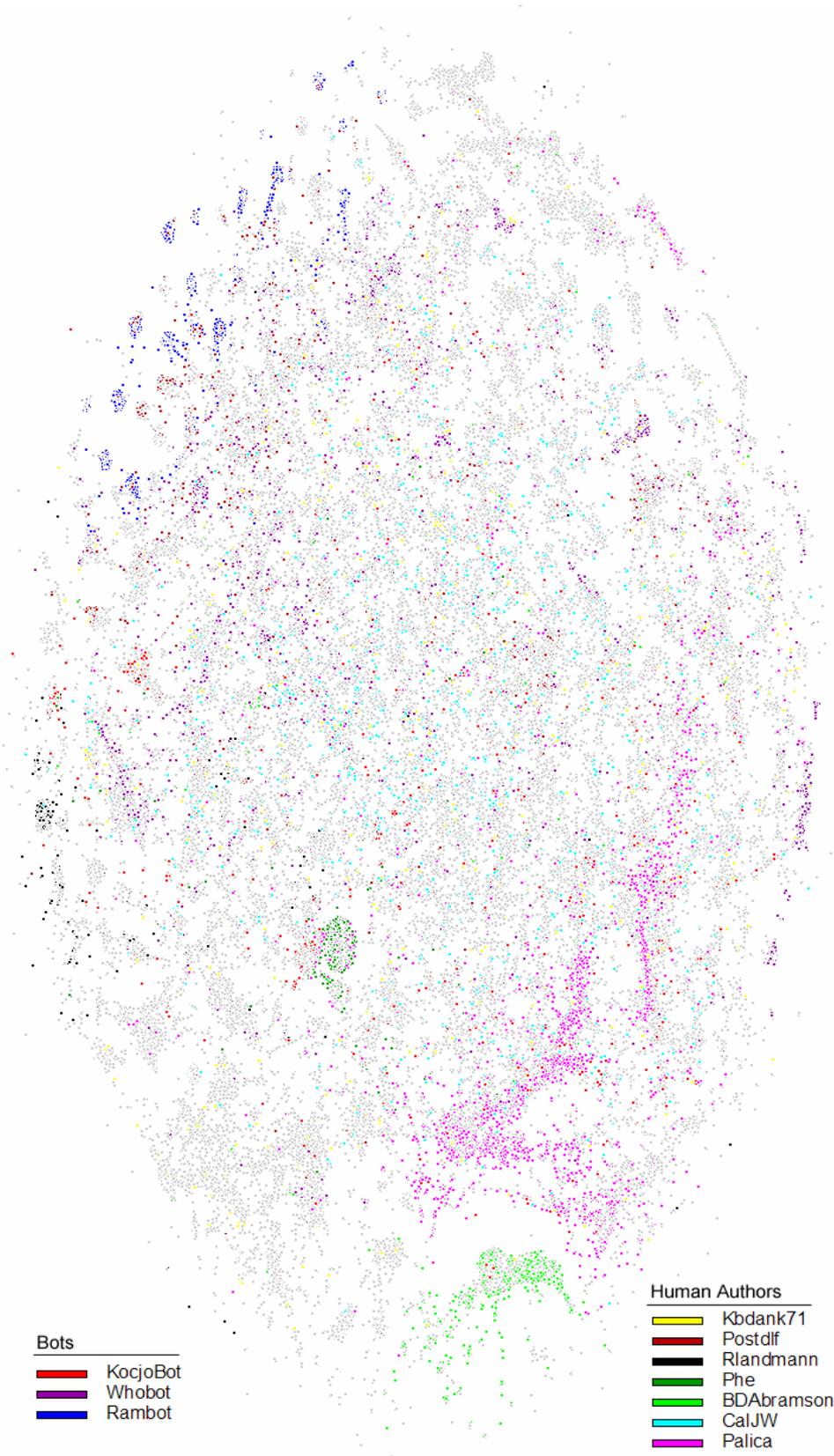

**Figure 6:** English Wikipedia category network colored by top ten most active authors.

## 5. Summary and Future Work

This paper presented, to our knowledge, the first semantic map of the English Wikipedia data. The map shows that when co-occurrence of categories within individual articles is considered as a measure of category similarity, categories appear as nice, logical clusters. The map also reveals that the category structure is well-maintained, although the bots and users involved in its maintenance are involved with varied scope and intentions.

The semantic map was created using only the data in the 'cur' and 'categorylinks' tables. We plan to continue this work by creating similar maps using historical versions of Wikipedia data in the 'old' table to study the evolution of Wikipedia. We will be looking closely at which clusters are most active at different periods, and what the catalysts for the different activities are. We will consider how the category structure has affected the evolution of Wikipedia, and develop novel methods for analyzing and visualizing this semi-hierarchical structure.

Further plans also include mapping the semantic structure of not just the categories, but also the articles in Wikipedia. Additionally we intend to examine other language Wikipedias to discover differences and similarities in the evolution, category structure, communities, and catalysts for change in Wikipedia.

**Acknowledgements**

We would like to thank Gavin LaRowe and Elijah Wright for insightful comments, technical advice, and administrative support. This work did benefit from discussions with Deborah MacPherson and Kevin W. Boyack. We would like to thank top Wikipedia contributors Brian Dean Abramson, Kris Dankovits, Klemen Kocjancic, Ruediger Landmann, Derek Ramsey, and 'Who' for their comments and feedback.

We acknowledge our collaboration with the Wikipedia Networks Team, in particular the frequent discussions with Vinko Zlatić, who provided the original inspiration for this collaboration, and valuable input from Hrvoje Štefančić, Mladen Domazet and Paul Stubbs. The Wikipedia Networks Team (http://www.idd.hr/en/wiki/WikiNet_Team) is an interdisciplinary research collaboration based in Zagreb, Croatia, studying the structure and dynamics of complex networks of article interlinkage across different language versions of Wikipedia. Its members are: Vinko Zlatić and Hrvoje Štefančić, Institute Ruđer Bošković, Zagreb, Miran Božičević, Multimedia Institute, Zagreb and Institute for Social Research in Zagreb, Mladen Domazet and Filip Miličević, Institute for Social Research in Zagreb, and Paul Stubbs, The Institute of Economics, Zagreb.

We thank the Wikimedia Foundation Inc. for making data dumps of Wikipedia freely available.

This work is supported by a National Science Foundation under IIS-0513650 and a CAREER Grant IIS-0238261 and a James S. McDonnell Foundation grant in the area Studying Complex Systems to Börner.